\documentclass[aps,pra,notitlepage,twocolumn,showpacs,superscriptaddress]{revtex4-1}  
\usepackage{blindtext}
\usepackage{graphicx}  
\usepackage{dcolumn}   
\usepackage{amsmath,bm,amssymb,latexsym,galois,amsthm}   
\usepackage{dsfont} 
\usepackage{mathrsfs}
\usepackage[all,cmtip]{xy}

\theoremstyle{plain}
\theoremstyle{remark}

\newtheorem{exmp}{Example}
\newtheorem{prop}{Proposition}

\newtheorem*{pf}{Proof}
\usepackage[unicode=true,bookmarks=true,bookmarksnumbered=false,bookmarksopen=false,breaklinks=false,pdfborder={0 0 1},backref=false,colorlinks=true]{hyperref}
\hypersetup{linkcolor=magenta, urlcolor=blue, citecolor=blue, pdfstartview={FitH}, hyperfootnotes=false, unicode=true}

\makeatletter

\begin{document}

\widetext

\title{Characterizing Nonlocal Correlations via Universal Uncertainty Relations}

\author{Zhih-Ahn Jia}
\email{Email address: giannjia@foxmail.com}
\affiliation{Key Laboratory of Quantum Information, Chinese Academy of Sciences, School of Physics, University of Science and Technology of China, Hefei, Anhui, 230026, P. R. China}
\affiliation{Synergetic Innovation Center of Quantum Information and Quantum Physics, University of Science and Technology of China, Hefei, Anhui, 230026, P. R. China}
\author{Yu-Chun Wu}
\email{Email address: wuyuchun@ustc.edu.cn}
\affiliation{Key Laboratory of Quantum Information, Chinese Academy of Sciences, School of Physics, University of Science and Technology of China, Hefei, Anhui, 230026, P. R. China}
\affiliation{Synergetic Innovation Center of Quantum Information and Quantum Physics, University of Science and Technology of China, Hefei, Anhui, 230026, P. R. China}
\author{Guang-Can Guo}
\affiliation{Key Laboratory of Quantum Information, Chinese Academy of Sciences, School of Physics, University of Science and Technology of China, Hefei, Anhui, 230026, P. R. China}
\affiliation{Synergetic Innovation Center of Quantum Information and Quantum Physics, University of Science and Technology of China, Hefei, Anhui, 230026, P. R. China}

\date{\today}

\begin{abstract}
Characterization of nonlocal correlations is one of the most attractive topics in quantum information theory. In this work, we develop the methods of detecting entanglement and steering based on the universal uncertainty relations and the fine-grained uncertainty relations. According to majorization form of the uncertainty relations, the uncertainty quantifier can be constructed by choosing Schur concave functions. Hence a large number of quantifier-independent entanglement and steering criteria are derived, from which many existing criteria based on different quantifiers can be rededuced. Finally, we show that entanglement and steering of pure states and some mixed states such as isotropic states can be always witnessed by some uncertainty quantifier.
\end{abstract}

\maketitle

\section{Introduction}

In quantum information theory, to distinguish and verify three kinds of nonlocal correlations~\cite{bell1987einsteinpodolskyrosen,bell1966,einstein1935,schrodinger1935discussion,Horodecki2009,Reid2009,Brunner2014bell}: entanglement, Einstein-Podolsky-rosed (EPR) steering and Bell nonlocality is one of the most fundamental problems. Not only do those three correlations tell the differences between quantum and classic worlds in different level theoretically, but also they showed the utilitarian power in quantum information processing tasks \cite{Horodecki2009,Reid2009,Brunner2014bell}. Although many progresses have been made, to answer whether a quantum state (known or unknown) is $\mathcal{C}$-correlated ($\mathcal{C}$=entanglement, steering, nonlocality) or not is still an open question \cite{Guhne2009,Reid2009,Brunner2014bell,Cavalcanti2017}. Experimentally, one expects to detect $\mathcal{C}$-correlation without knowing the state completely because of the high complexity of state tomography. How can we determine if the state is $\mathcal{C}$-correlated or not with some finite measurement settings is a very important task.

For entanglement verification, we assume the precise knowledge of the dimensions of the systems and the operations chosen to implement, but for many information tasks, this is too stringent. Motivated by device-independent quantum information protocols for which minimal assumptions are desired, fully device-independent entanglement detection method are also developed, which are based on the experimentally very demanding observation of the violation of Bell inequalities without detection loopholes~\cite{hensen2015loophole,Shalm2015,Giustina2015}. EPR steering was recognised as a special kind of nonlocal correlations which is intermediate between entanglement and Bell nonlocality, and it is characterized by the failure of local hidden states (LHS) model. Witnessing steering implies entanglement in bipartite systems without any assumption of one of the parties, this kind of one-side device-independent detection is less stringent than that of the violation of Bell inequalities and it is much more feasible in experiments~\cite{wittmann2012loophole} and is more robust against experimental noise~\cite{saunders2010experimental,handchen2012observation,de2012conclusive,fuwa2015experimental,Bennet2012}. Steering was initially introduced by Einstein \emph{et al.}~\cite{einstein1935} and Schr\"{o}dinger~\cite{schrodinger1935discussion} and recently rigourously defined by Wiesman \emph{et al.} to describe the ability of one experimenter Alice to remotely prepare the ensemble of states for another experimenter Bob by performing a local measurement on her half of the sharing state and communicating the measurement results to Bob~\cite{Wiseman2007}.  Steering also has been found useful in a large number of quantum information tasks, such as one-side device-independent quantum key distribution~\cite{Tomamichel2011,Beanciard2012}, one-side device-independent estimation of measurement incompatibility~\cite{Cavalcanti2016a}, subchannel discrimination~\cite{Piani2015}, one-side device-independent self-testing~\cite{Supic2016,Gheorghiu2017} one-sided device-independent quantum secret sharing \cite{armstrong2014multipartite} and secure quantum teleportation \cite{He2015}.

Compared with the entanglement and Bell nonlocality, EPR steering's characterization is relatively incomplete and less studied. With the increasing interests on this phenomenon recent years, researchers have developed many steering criteria, such as linear inequality criterion~\cite{Cavalcanti2009,Pusey2013}, criterion based on conditional standard deviation~\cite{Reid1989,Cavalcanti2009}, entropy~\cite{Walborn2011,Schneeloch2013}, and fine-grained uncertainty relations (URs)~\cite{Pramanik2014}. These criteria are more or less inspired by the seminal work of Ried~\cite{Reid1989}, which strongly depends on the Heisenberg's UR~\cite{Heisenberg1927}.  It is worth mentioning that entanglement detection method based on URs with variances and entropies of measurements has also been developed~\cite{guhne2004,Horodecki2009}, but there is no similar result for Bell nonlocality detection up to now as far as we know.

In this work, we develop the criteria of entanglement and steering based on the quantifier-independent formulation of UR (we focus on the so-called universal UR, which is a special case of quantifier-independent UR but already very general), and we also develop the fine-grained UR based criteria, which are the complementary part of the detection based on universal UR. Almost all existing UR-based criteria (e.g. based on standard deviation or entropy) can be subsumed within this universal detection method.

The pager is organized as follows. After introducing the quantifier-independent UR in particular, universal UR in Sec. \ref{sec:UR}, we develop the entanglment and steering detecting methods based on universal UR in Sec. \ref{sec:Ent}. In Sec. \ref{sec:Fine-UR}, we develop the detection method base on fine-grained URs and we discuss the differences and similarities between fine-grained UR based method and universal UR based method. Finally, we conclude and explore future directions of the study.

\section{Quantifier-independent UR}
\label{sec:UR}

The UR is a fundamental intrinsic limitation in the description of the quantum world, it is the most revolutionary viewpoint of our understanding of the nature which is sharply different with our intuition from everyday life. It was Heisenberg~\cite{Heisenberg1927} who first revealed the intrinsic impossibility of joint predictability for position and momentum operators. He showed that the lower standard deviation of position leads a larger standard deviation of momentum and vice versa, which means that we can not simultaneously detect both position and momentum with precise value. In later 1929, Robertson~\cite{Robertson1929} extended the relation to any pair of observables
\begin{equation}\label{HUR}
\Delta(x) \Delta(y) \geq \frac{1}{2}|\langle \Psi|[x,y]|\Psi\rangle |.
\end{equation}
Although this is a great triumph in our understanding of quantum mechanics, there are still some shortcomings of the above Heisenberg-Robertson UR, one is that the lower bound of Eq. (\ref{HUR}) may be zero even if $x$ and $y$ are not commutative, and the other is that the left hand side of Eq. (\ref{HUR}) may change with mere relabeling of the outcomes~\cite{Deutsch1983}.

In general, UR is in fact a trade-off relation of the spreads of several probability distributions. For example, in the $d$-dimensional probability space, the distribution $\mathbf{e}_j=(0,\cdots,0,1,0,\cdots,0)$ is the most certain one, and the uniform distribution $\mathbf{p}=(1/d,\cdots,1/d)$ is the most uncertain one.  With the viewpoint that uncertainty of a distribution can not decrease by relabeling of the outcomes, Deutsch~\cite{Deutsch1983} argued that standard deviation can not be used as a \emph{quantitative} description of UR. A modern method to characterize the URs is based on quantum information terms (e.g., entropy), the most typical entopic UR was initiated by Bia{\l}ynicki-Birula and Mycielski~\cite{bialynicki1975uncertainty} and later developed by Deutsch~\cite{Deutsch1983} for finite spectrum non-degenerate observables, then conjectured by Kraus~\cite{Kraus1987} and later proved by Maassen and Uffink~\cite{Maassen1988} for general non-degenerate observables $x$ and $y$ which sharing no eigenstates, here we give the the relation for any state (pure or mixed)
\begin{equation}\label{EUR}
H(x)_{\rho}+H(y)_{\rho}\geq -\log_2 c_{xy}+H(\rho),
\end{equation}
where $c_{xy}=\mathrm{max}_{i,j}|\langle \phi_i| \psi_j\rangle |^2$ is the maximum overlap between the eigenstates $|\phi_i\rangle$ and $|\psi_j\rangle$ of $x$ and $y$, which quantifies the complementarity of $x$ and $y$. By traversing over all states we get the usual Maassen-Uffink form $H(x)+H(y)\geq -\log_2 c_{xy}$. Note that the assumption that $x$ and $y$ share no common eigenstates can also been taken out by modifying the lower bound of Eq.(\ref{EUR}), see Ref.~\cite{krishna2002entropic}.

The entropic UR overcomes some shortcomings of Heisenberg-Robertson UR, but there is no reason that entropy function is the most adequate uncertainty quantifier. In 2011, Partovi~\cite{Partovi2011} proposed a new formulation of UR based on majorization theory, which was later improved by Friedland \emph{et al.}~\cite{Friedland2013} and Pucha{\l}a \emph{et al.}~\cite{Puchala2013} independently, this kind of UR which we refer to as universal UR are not limited to considering only entropic functions, but also any nonnegative Schur-concave functions. There, the only assumption for an uncertainty quantifier is that uncertainty of a probability distribution cannot decrease under so called random relabeling~\cite{Friedland2013,poostindouz2016fine}, which means that if $\Omega$ is an uncertainty quantifier, then for a probability distribution $\mathbf{p}$ and its random relabeling $\mathbf{q}=\sum_{\pi \in \mathfrak{S}_d} p(\pi) R_{\pi} \mathbf{p}$ (where $\mathfrak{S}_d$ is the permutation group of order $d$ and $R_{\pi}$ is the $d$-dimensional matrix representation of permutation $\pi$), it must satisfy $\Omega(\mathbf{p})\leq \Omega(\mathbf{q})$.

To formulate the UR in a more unified way, we need the notion of probability vectors and doubly stochastic matrix. By $\mathbf{p}=(p_1,\cdots,p_d)$ a probability vector we means that $\mathbf{p}$ is a real vector with each entry non-negative and $\sum_{i} p_{i}=1$; while a doubly stochastic matrix $D_{ij}$ is a real matrix with each entry non-negative and $\sum_i D_{ij}=1$ for all $j$ and vice versa. Note that tensor product and convex combination of two probability vectors (resp. doubly stochastic matrice) is still a probability vector (resp. doubly stochastic matrix). Hereinafter we assume that every probability vector is of $d$-dimension, since we can add trailing zeros whenever necessary. And we use $\mathbf{p}^{\uparrow}$(resp. $\mathbf{p}^{\downarrow}$) to denote the rearranged version of $\mathbf{p}$ in a non-decreasing (resp. non-increasing) order. We define the $i$-th up-going function (resp. $j$-th down-going function) of a probability vector as $f_i^{\uparrow}(\mathbf{p})= p_i^{\uparrow}$ (resp. $g_j^{\downarrow}(\mathbf{p})=p_j^{\downarrow}$), where $p_i^{\uparrow}$ (resp. $p_j^{\downarrow}$) is the $i$-th (resp. $j$-th) component of $\mathbf{p}^{\uparrow}$ (resp. $\mathbf{p}^{\downarrow}$). A vector $\mathbf{p}$ is majorized by a vector $\mathbf{q}$, denoted as $\mathbf{p}\preceq \mathbf{q}$, whenever $\sum_{i=1}^k p^{\downarrow}_i\leq \sum_{i=1}^k q^{\downarrow}_i$, for all $1\leq k < d$. By Birkhoff theorem~\cite{Birkhoff1946,bhatia1997matrix}, which states that the probabilistic mixture of permutation matrices is a doubly stochastic matrix, we have $\mathbf{p}\preceq \mathbf{q}$ if and only there is a doubly stochastic matrix $D$ such that $\mathbf{p}= D\mathbf{q}$.

Suppose that we can prepare a large number of identical state $\rho$, then by repeating the experiment many times we can collect the probability distributions of a set of measurements $\mathbf{x}=\{x_i\}_{i=1}^m$ with corresponding positive operator-valued measurements(POVM) $\{F_{a_i|x_i}\}_{a_i=1}^d$ as probability vectors $\mathbf{p}_{x_i}(\rho)=(\mathrm{tr}(F_{1|x_i}\rho),\cdots,\mathrm{tr}(F_{d|x_i}\rho))$. The object of investigation of UR is therefore the joint probability distribution $\mathbf{p}_{x_1}(\rho)\otimes \cdots \otimes \mathbf{p}_{x_m}(\rho)$. As has been indicated in Refs.~\cite{Partovi2011,Friedland2013,Puchala2013}, the UR for a set of measurements $\mathbf{x}$ is of the form
\begin{equation}\label{}
\mathbf{p}_{x_1}(\rho)\otimes \cdots \otimes \mathbf{p}_{x_m}(\rho)\preceq \bm{\omega}_{\mathbf{x}}, \forall \rho \in \textsf{S}(\mathcal{H}),
\end{equation}
where the uncertainty bound vector $\bm{\omega}_{\mathbf{x}}$ is some probability vector which only depends on the chosen set of measurements $\mathbf{x}$, in particular, $\bm{\omega}_{\mathbf{x}}$ is independent of $\rho$, $\textsf{S}(\mathcal{H})$ is the set of all density operators on Hilbert space $\mathcal{H}$. In Ref.~\cite{Friedland2013}, Friedland \emph{et al.} derived an explicit expression of the upper bound of $\bm{\omega}_{\mathbf{x}}$ in general, especially in two dichotomic measurement case, they gave an exact expression for two measurements $\mathbf{x}=\{x_1, x_2\}$ as $\bm{\omega}_{\mathbf{x}}=(\gamma_1, \gamma_2-\gamma_1,0,0)$ where $\gamma_1=(1+c)^2/4$, $\gamma_2=(1+c')^2/4$ with $c=\max_{k,j}|\langle \phi_k|\psi_j\rangle|$ and $c'=\max_{k=k',j\neq j';k\neq k',j= j'}\sqrt{|\langle \phi_k|\psi_j\rangle|^2 +|\langle \phi_{k'}|\psi_{j'}\rangle|^2}$, $\phi_k$ and $\psi_j$ are eigenvectors of $x_1$ and $x_2$ respectively. Notice that `$\preceq$' is a partial order on the space of all probability vectors of $d$-dimension \cite{partialorder}, we cannot always compare any two distributions. Nonetheless, this formulation of UR is still very powerful, we can compare most distributions without using the more elusive uncertainty quantifier approach, this is also the reason why we use the terminology\textendash quantifier-independent UR\textendash in this work.

If we choose uncertainty quantifier as some  Schur-concave function $\Omega$ (see Refs.~\cite{bhatia1997matrix,marshall2011inequalities} for detailed discussions of Schur concave function and majorization theory), then the universal UR reads
\begin{equation}\label{}
\Omega(\mathbf{p}_{x_1}(\rho)\otimes \cdots \otimes \mathbf{p}_{x_m}(\rho))\geq \Omega(\bm{\omega}_{\mathbf{x}}),\forall \rho \in \textsf{S}(\mathcal{H}).
\end{equation}
Further, if $\Omega$ is additive under tensor products, then we have
\begin{equation}\label{eq:sum-UR}
 \Omega(\mathbf{p}_{x_1}(\rho))+\cdots+\Omega(\mathbf{p}_{x_m}(\rho))\geq \Omega(\bm{\omega}_{\mathbf{x}}), \forall \rho \in \textsf{S}(\mathcal{H}).
\end{equation}

We will use Eq.(\ref{eq:sum-UR}) to develop the criteria of entanglement and steering, since $\Omega$ is only restricted to be Schur concave function, many entropic UR is just some special case of universal UR and since Heisenberg-Robertson UR can be derived from entropic UR~\cite{Coles2014}, it can also be subsumed within this general framework.

\begin{figure}
\includegraphics[scale=0.30]{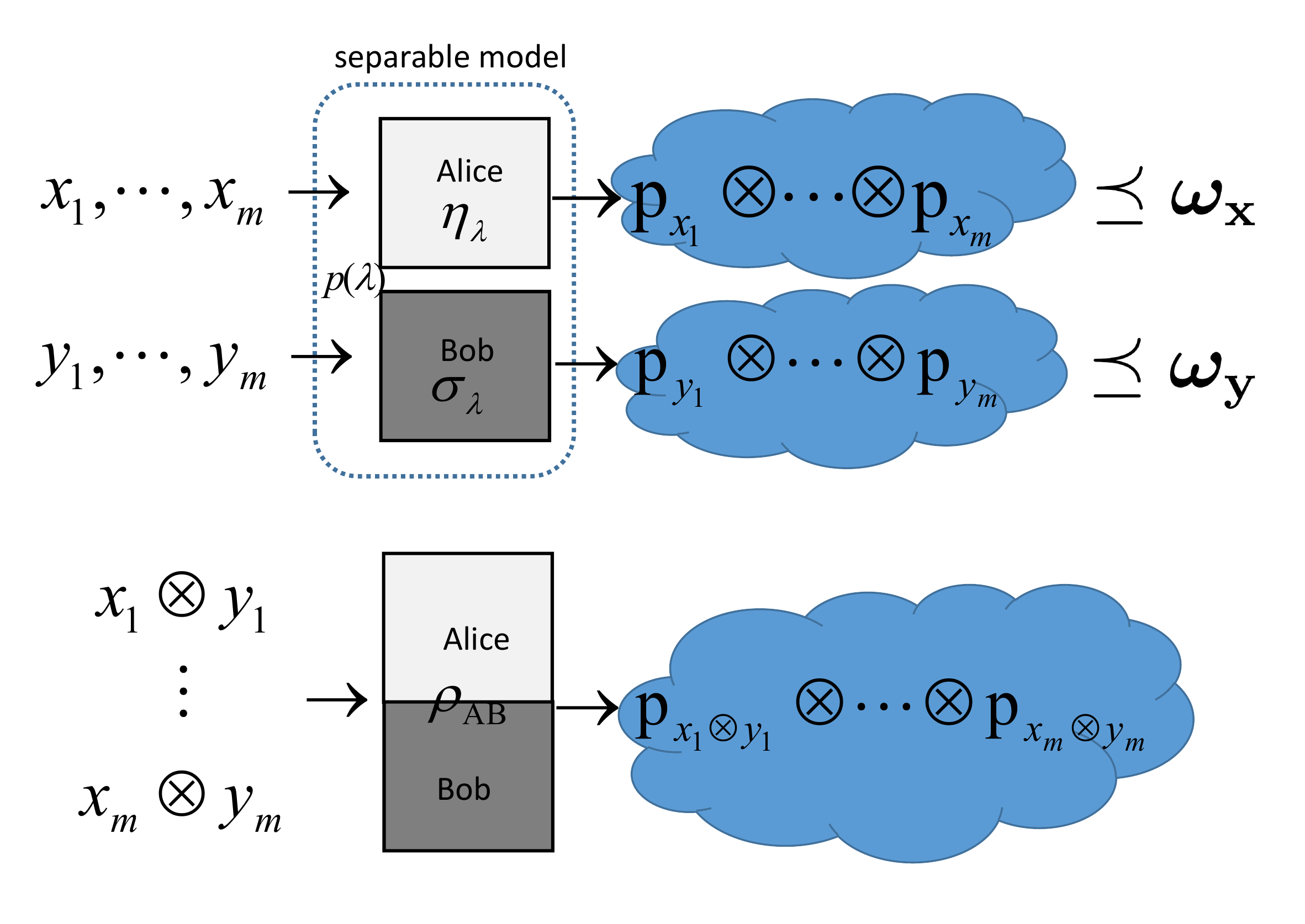}
\caption{\label{fig:ent}(color online). The sketch of the entanglement verification based on universal UR. Alice and Bob implement $x_i$ and $y_i$ in each run of experiment, then by repeating the experiment many times, they can collection the statistics $\mathbf{p}_{x_1\otimes y_1}\otimes \cdots \otimes \mathbf{p}_{x_m\otimes y_m}$, using this they can determine if their sharing state is entangled or not.}
\end{figure}

\section{Entanglement and steering verification based on universal UR}
\label{sec:Ent}
\subsection{Entanglement verification based on universal UR}

Consider two parties, Alice and Bob, each holding one half of a sharing state $\rho\in \textsf{S}(\mathcal{H}_A\otimes \mathcal{H}_B)$, where $\mathcal{H}_A$ and $\mathcal{H}_B$ are the Hilbert spaces of Alice and Bob respectively. To determine whether the shared state admits a separable model, i.e., $\rho=\sum_{\lambda}p(\lambda)\eta_{\lambda} \otimes \sigma_{\lambda}$ for some $\eta_{\lambda}\in \textsf{S}(\mathcal{H}_A)$ and $\sigma_{\lambda}\in \textsf{S}(\mathcal{H}_B)$ and a probability distribution $p(\lambda)$, Alice and Bob choose to implement some $d$-outcome measurements $\mathbf{x}=\{x_i\}_{i=1}^m$ and $\mathbf{y}=\{y_i\}_{i=1}^m$ respectively. We refer to this kind of detection scenario as $2$-party $m$-measurement $d$-outcome ($2$-$m$-$d$) entanglement scenario. Using the universal UR, we give the following entanglement criterion.

\begin{prop}
Let $\rho$ be a bipartite state shared by Alice and Bob which admits a separable model, Alice and Bob choose to implement measurements $\mathbf{x}=\{x_1,\cdots,x_m\}$ and $\mathbf{y}=\{y_1,\cdots,y_m\}$ respectively, then for any convex and Schur concave function $\Omega$, we have
\begin{equation}\label{eq:ent}
\sum_{i=1}^m\Omega(\mathbf{p}_{x_i\otimes y_i})\geq \max \{\Omega(\bm{\omega}_{\mathbf{x}}),\Omega(\bm{\omega}_{\mathbf{y}})\},
\end{equation}
where $\bm \omega_{\mathbf{x}}$ and $\bm \omega_{\mathbf{y}}$ are two respective uncertainty bound vectors of measurements $\mathbf{x}$ and $\mathbf{y}$. See Fig.(\ref{fig:ent}) for the sketch of the detection method.
\begin{pf}
First of all, we observe that for product state $\rho=\eta\otimes \sigma$ and product measurement $x\otimes y$, we have $\mathbf{p}_{x\otimes y}(\rho) \preceq \mathbf{p}_{x}(\eta)$ and $\mathbf{p}_{x\otimes y}(\rho) \preceq \mathbf{p}_{y}(\sigma)$. See Ref.~\cite{Guhne2004a} for the proof of this observation. Then suppose that the sharing state $\rho$ admits a separable model, i.e., $\rho=\sum_{\lambda}p(\lambda)\eta_{\lambda} \otimes \sigma_{\lambda}$, for each product measurement $x_i\otimes y_i$, the measurement statistics $\mathbf{p}_{x_i\otimes y_i}$ is just the convex combination $\sum_{\lambda}p_{\lambda} \mathbf{p}_{x_i\otimes y_i}(\eta_{\lambda}\otimes \sigma_{\lambda})$. By the convexity of $\Omega$, we get $\Omega(\mathbf{p}_{x_i\otimes y_i}(\rho))\geq \sum_{\lambda}p_{\lambda}\Omega(\mathbf{p}_{x_i\otimes y_i}(\eta_{\lambda}\otimes \sigma_{\lambda}))$.  Summing over $i$, and using the first observation we get $\sum_{i=1}^m\Omega(\mathbf{p}_{x_i\otimes y_i})\geq \max \{\sum_{\lambda} \sum _{i}p(\lambda) p_{x_i}(\eta_{\lambda}),\sum_{\lambda} \sum _{i}p(\lambda) p_{y_i}(\sigma_{\lambda})\}$. Then using the universal UR $\sum_i\Omega(\mathbf{p}_{x_i}(\eta_{\lambda}))\geq \Omega(\bm{\omega}_{\mathbf{x}})$ and $\sum_i\Omega(\mathbf{p}_{y_i}(\sigma_{\lambda}))\geq \Omega(\bm{\omega}_{\mathbf{y}})$ for all $\lambda$ we arrive at the conclusion.
\qed
\end{pf}
\end{prop}

Intuitively, Eq.(\ref{eq:ent}) means that the distributions of measurements globally implemented to the entangled states are not more uncertain than the one implemented locally on separable states. And the convexity condition  we made for uncertainty quantifier $\Omega$ has a clear physical meaning: one can not decrease uncertainty of several probability vectors by probabilistically mixing them together. There are many such kind of functions, e.g., Shannon entropy, minimum entropy and so on. Notice that here we concentrate on the case that Alice and Bob choose the same number of measurements and each measurement results in one of $d$ outcomes for clarity, but our result can also be extended into the more complicated case, e.g., the numbers of measurements in $\mathbf{x}$ and $\mathbf{y}$ are different \cite{ent}.

For pure entangled state, Eq.(\ref{eq:ent}) can get a violated value. Suppose that $|\psi\rangle$ is a pure entangled state which is also the eigenstate of $\{x_i\otimes y_i\}_{i=1}^m$, then $\mathbf{p}_{x_1\otimes y_1}\otimes \cdots \otimes \mathbf{p}_{x_m\otimes y_m}$ will be the most certain probability vector $(0,\cdots,0,1,0,\cdots,0)$, thus $\mathbf{p}_{x_1\otimes y_1}\otimes \cdots \otimes \mathbf{p}_{x_m\otimes y_m} \succeq \bm{\omega}_{\mathbf{x}},\bm{\omega}_{\mathbf{y}}$ , the inequality can be violated. Actually, by rewriting each pure entangled state in its Schmidt decomposition form and choosing the product measurements as proper Gellman matrices in the basis containing these Schmidt vectors, each pure entangled states can be witnessed with this criterion. Here we choose Shannon entropy to give an straightforward exemplary illustration of the power of the criterion.

\begin{exmp}\label{em:ent}
Suppose that Alice and Bob sharing a bipartite state $\rho$, for the measurements $\{x,y\}$,  by choosing Shannon entropy as the uncertainty quantifier, we have
\begin{equation}\label{}
H(x^Ax^B)+H(y^Ay^B)\geq H(\bm{\omega}_{xy}).
\end{equation}
For two-qubit state, by choosing to measure Pauli measurements $\sigma_x$ and $\sigma_y$, from calculation we know that $\bm{\omega}_{\sigma_x\sigma_y}=(\frac{3+2\sqrt{2}}{8},\frac{5-2\sqrt{2}}{8},0,0)$, then we have
$H(\sigma_x^A\sigma_x^B)+H(\sigma_y^A\sigma_y^B)\geq  H(\bm{\omega}_{\sigma_x\sigma_y})\simeq 0.844$, violation of the inequality is a signature of the entanglement. Note that this  bound is less that the bound $1$ given by Giovannetti~\cite{Giovannetti2004} numerically and by G\"{u}hne and Lewenstein~\cite{Guhne2004a} analytically for Shannon entropy based separability criterion, this is because $\bm{\omega}_{\sigma_x\sigma_y}$ is a quantifier-independent quantity, we can choose other quantifier to remedy the shortage in Shannon type inequality. For example, by choosing minimum entropy $H_{\infty}(\mathbf{p})$, it can remedy the shortage of Shannon entropy in detecting the entanglement of Werner states, this has been show in the entropy UR based case~\cite{Guhne2004a}, this is a strong evidence for the power of quantifier-independent criterion.
\end{exmp}

\subsection{Steering detection based on universal UR}

Consider two distant parties, Alice and Bob, sharing a state $\rho$ with reduced states $\rho_A$ and $\rho_B$ respectively, Alice can choose $m$ different measurement settings denoted as $\mathbf{x}'=\{x'_1,\cdots, x'_m\}$, each of measurement results in one of $k$ outcomes ${\mathbf{a'}}_{x'_i}=(1_{x'_i},\cdots, k_{x'_i})$. For steering test, we have the knowledge of local Hilbert space of $\mathcal{H}_B$ of Bob's side, and the space dimension is $d=\dim (\mathcal{H}_B)$. Bob's available data are the assemblage of steered state $\mathscr{A}_{\mathbf{x'}}=\{\{\sigma_{a'_i|x'_i}\}_{a'_i}\}_{x'_i}$ with $\sigma_{a'_i|x'_i}\geq 0$, $\mathrm{tr}(\sigma_{a'_i|x'_i})=p(x'_i=a'_i)$ and $\sum_{a'_i}\sigma_{a'_i|x'_i}=\rho_B$. We call the assemblage $\mathscr{A}_{LHS}=\{p(\lambda),\sigma_{\lambda}\}_{\lambda\in \Lambda}$ where $p(\lambda)\geq 0$, $\sum_{\lambda}p(\lambda)=1$, the density matrix $\sigma_{\lambda}\geq 0$ and $\sum_{\lambda}p(\lambda)\sigma_{\lambda}=\rho_B$, LHS assemblage in $m$-$k$-$d$ steering scenario, if for any $m$ $k$-valued measurements $\mathbf{x}'$, we have $\sigma_{a'_i|x'_i}=\sum_{\lambda\in \Lambda}p(\lambda)p(x'_i=a'_i|\lambda)\sigma_{\lambda}$. The shared state is steerable if there is some measurement set $\mathbf{x}'$, such that the steered assemblage $\mathscr{A}_{\mathbf{x}'}$ does not admit a LHS model.

The intuition behind the steering inequality base on UR is what is so called Reid's extension of EPR's sufficient condition of reality~\cite{Cavalcanti2009}, which states that if we can predict the value of a physical observable with some specified uncertainty without disturbing the system, then there is a stochastic element of reality which determines the physical quantity with at most that specified uncertainty.

\begin{figure}
\includegraphics[scale=0.40]{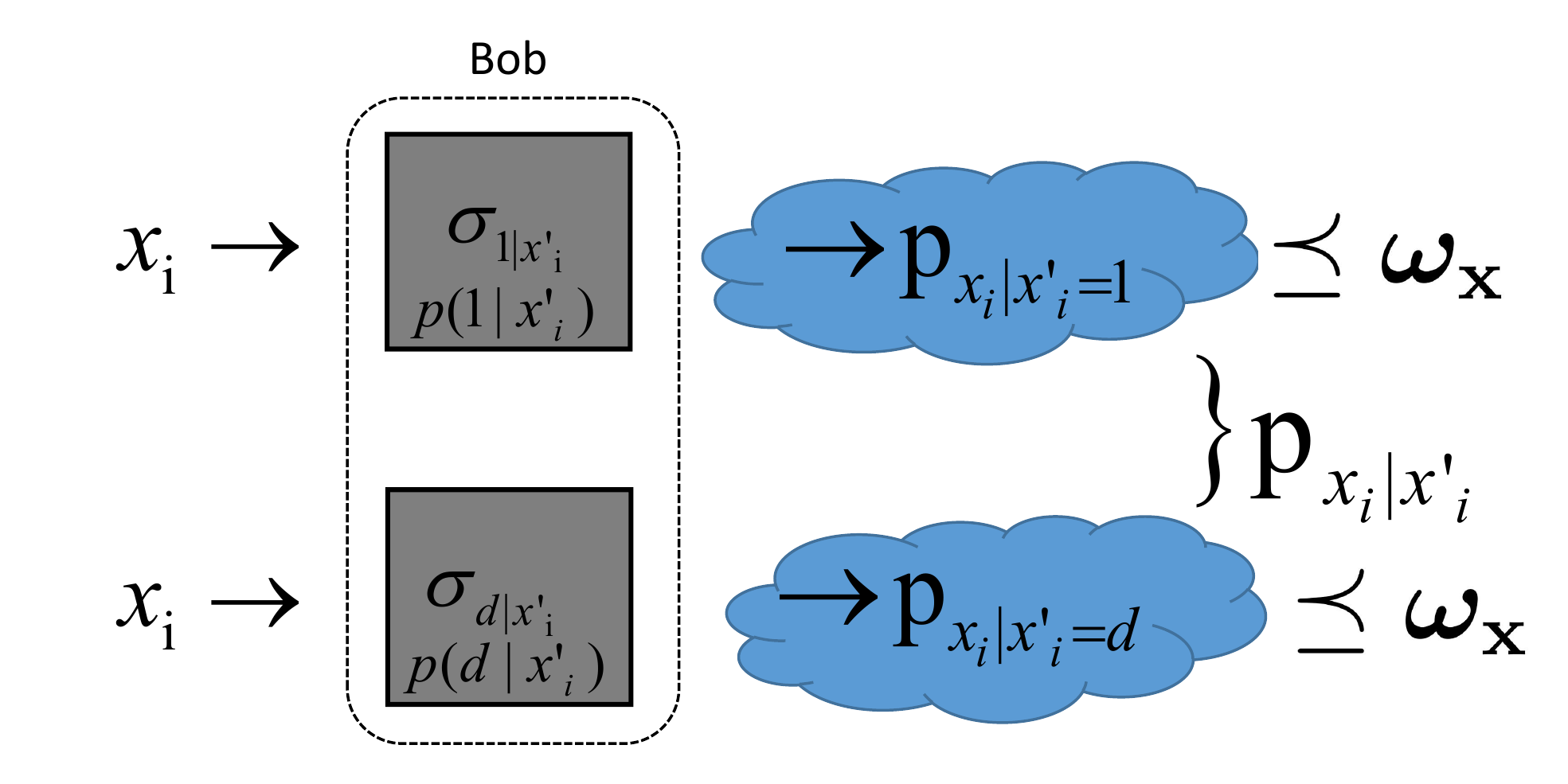}
\caption{\label{fig:steering}(color online). The sketch of the steering detection based on universal UR. Bob ask Alice to steer his state in the ensemble $\mathscr{A}_{x'_i}$, then he choose to measure $x_i$ on his states, after running all ensembles of assemblage $\mathscr{A}_{\mathbf{x}'}=\{\mathscr{A}_{x_i}\}_{x_i}$, he can determine if $\mathscr{A}_{\mathbf{x}'}$ is steerable or not, i.e., if shared state is steerable or not.}
\end{figure}

Now we are in a position to explain how to detect EPR steering using the universal UR. In the $m$-$d$-$d$ steering scenario, to test if the share two partite state $\rho$ is steerable or not, Bob ask Alice to steer his state in an assemblage of ensembles $\{\{\sigma_{a'_i|x'_i}\}_{a'_i}\}_{x'_i}$, Bob then choose to measure $m$ $d$-valued measurements $\mathbf{x}=\{x_i\}$ on each respective ensemble of his subsystem. After repeating the experiment many times all information he can collect from experiment is $d^2\times m$ probability distributions $\{p(a_i,a'_i|x_i,x'_i)\}$. With the constraint of non-signaling principle, he can also get the probability distributions of $\{p(a_i|x_i)\}$ and $p(a'_i|x')$ from marginal of the joint distributions. We use the conditional probability vector
 $\mathbf{p}_{x_i|x'_i=a'_i}$ to denote $(p(x_i=1|a'_i),\cdots, p(x_i=d|a'_i))$. Then we have the following proposition:

\begin{prop}\label{prop:test} Let $\Omega$ be an uncertainty quantifier, which we choose as a convex Schur concave function. If the state $\rho_{AB}$ admits LHS model for any measurement assemblage, then the following inequality holds,
\begin{equation}\label{eq:Steering test}
\sum_{i=1}^m\Omega(\mathbf{p}_{x_i|x'_i})\geq \Omega(\bm{\omega}_{\mathbf{x}}),
\end{equation}
where $\Omega(\mathbf{p}_{x_i|x'_i})=\sum_{a'_i}p(a'_i)\Omega (\mathbf{p}_{x_i|x'_i=a'_i})$, and $\bm{\omega}_{\mathbf{x}}$ is the uncertainty bound vector for measurement $\mathbf{x}$. See Fig.(\ref{fig:steering}) for the sketch of the detection method.
\begin{pf}
Firstly, note that we have UR for set of measurements $\mathbf{x}$ as $\sum_{i=1}^m\Omega(\mathbf{p}_{x_i}(\rho))\geq \Omega(\bm{\omega}_{\mathbf{x}})$ for all $\rho\in \textsf{S}(\mathcal{H}_B)$, and the conditional probability vector we collect from experiment is $\mathbf{p}_{x_i|x'_i=a'_i}=(p(x_i=1|x'_i=a'_i),\cdots,p(x_i=d|x'_i=a'_i))$, where $p(x_i=j|x'_i=a'_i)=\mathrm{tr}(F_{x_i=j}\sigma_{a'_i|x'_i})/p(a'_i|x'_i)$. If the sate $\rho_{AB}$ admits a LHS model $\{p(\lambda),\rho_{\lambda}\}$, then $\mathbf{p}_{x_i|x'_i=a'_i}=\sum_{\lambda}p(\lambda|x'_i=a'_i)\mathbf{p}_{x_i}(\rho_{\lambda})$, by taking the convex Schur concave uncertainty quantifier, we get $\Omega(\mathbf{p}_{x_i|x'_i=a'_i})\geq\sum_{\lambda}q(\lambda|x'_i=a'_i)\Omega(\mathbf{p}_{x_i}(\rho_{\lambda}))$. Then by averaging on the probability distribution $\{p(x'_i=a'_i)\}_{a'_i}$, we arrive at $\Omega(\mathbf{p}_{x_i|x'_i})\geq \sum_{\lambda}q(\lambda)\Omega(\mathbf{p}_{x_i}(\rho_{\lambda}))$. Summing over all $x_i$ and using the UR we can get Eq. (\ref{eq:Steering test}).\qed
\end{pf}
\end{prop}
Now let us apply the proposition \ref{prop:test} to steering tests.

\begin{exmp}
We still take Shannon entropy as uncertainty quantifier, then in $2$-$d$-$d$ steering scenario, we have the following steering criteria,
\begin{equation}\label{}
H(x^B|x^A)+H(y^B|y^A)\geq H(\bm{\omega}_{xy}).
\end{equation}
For two-qubit state, if we choose $x$ and $y$ as Pauli operator $\sigma_x$ and $\sigma_y$, then the lower bound of the inequality will still been about $0.844$, which is less compared with the usual entropy steering inequality with the lower bound $-\log_2 c_{\sigma_x\sigma_y}= 1$. As we have indicated in the example \ref{em:ent}, this shortage origins from the universality of $\bm{\omega}_{\sigma_x\sigma_y}$ and can be overcame by choosing other uncertainty quantifier.
\end{exmp}

\section{Fine-grained UR based detecting methods.}
\label{sec:Fine-UR}
\subsection{Detecting methods}
In some respects, the UR based on standard deviation and entropy is incomplete, since they only account for the uncertainty among several distributions with regarding each distribution as a whole. They can not be used to distinguish the uncertainty inherent in obtaining a particular combination of outcomes of several measurements. For this reason, Oppenheim and Wehner~\cite{Oppenheim2010} introduced the so called fine-grained UR. At the same year, Berta \emph{et al.}~\cite{Berta2010uncertainty} extended the entropic UR into the memory-assist scenario, but extending universal UR and fine-grained UR to this scenario is nontrivial and remains open. Here, as a complements of the detecting methods based on universal UR, we propose the detecting methods based on fine-grained UR.

For a given state $\rho$, if we choose $m$ measurements $\mathbf{x}=(x_1, \cdots ,x_m)$ with \emph{a priori} probabilities $p(\mathbf{x})=\{p(x_i)\}_{i=1}^m$ and each measurement results in $d$ possible outcomes $a_x\in \mathbb{Z}_d=\{0, \cdots ,d-1\}$, then for each possible outcome string $\mathbf{a}=(a_1,\cdots,a_m)\in (\mathbb{Z}_d)^{m}$, there are some constraint of these statistics restricted by uncertainty principle:
\begin{equation}\label{uncertainty}
\sum_{i=1}^{n}p(x_i)p(a_i|x_i)_{\rho}= B_{\mathbf{a}}^{\mathbf{x}}(\rho)\leq B_{\mathbf{a}}^{\mathbf{x}}, \forall \rho \in \textsf{S}(\mathcal{H}),
\end{equation}
where the upper bound is defined as $B_{\mathbf{a}}^{\mathbf{x}}:=\mathrm{max}_{\rho} B_{\mathbf{a}}^{\mathbf{x}}(\rho)$, the maximization is taken over all quantum states of some physical system. Notice that there are $d^m$ constraints in total for a set of given observables $\mathbf{x}$, we call these inequalities the fine-grained UR of $\mathbf{x}$.

We first give the entanglement criterion base on fine-grained UR, similar results have been developed in Ref.~\cite{Rastegin2016}.

\begin{prop}\label{prop:f-e}
If Alice and Bob share a state $\rho$ which admits a separable model, then for any observables $\mathbf{x}=\{x_i\}_{i=1}^m$ and $\mathbf{y}=\{y_i\}_{i=1}^m$ chosen by Alice and Bob respectively, and for all outcome strings $\mathbf{a}=(x_1=a_1,\cdots,x_m=a_m)$ of $\mathbf{x}$ and $\mathbf{b}=(y_1=b_1,\cdots,y_m=b_m)$ of $\mathbf{y}$ , we have
\begin{equation}\label{eq:f-ent}
\sum_{i,j=1}^m p(x_iy_j)p(a_ib_j)\leq B^{\mathbf{x}\mathbf{y}}_{{\mathbf{a}\mathbf{b}}}.
\end{equation}
where $B^{\mathbf{x}\mathbf{y}}_{{\mathbf{a}\mathbf{b}}}$ is uncertainty bound for outcome string $\mathbf{a}\otimes \mathbf{b}$ under \emph{a priori} distribution $p(x_iy_j)$, and it is obtained by taking maximum over all product states, i.e. $B^{\mathbf{x}\mathbf{y}}_{{\mathbf{a}\mathbf{b}}}=\max B^{\mathbf{x}\mathbf{y}}_{{\mathbf{a}\mathbf{b}}}(\eta\otimes \sigma)$ with $\eta \in \textsf{S}(\mathcal{H}_A)$ and $\sigma \in \textsf{S}(\mathcal{H}_B)$.
\begin{pf}
Suppose that $\rho=\sum_{\lambda}p_{\lambda}\eta_{\lambda}\otimes \sigma_{\lambda}$, with the observation that $\sum_{i,j=1}^m p(x_iy_j)p(a_ib_j)_{\eta_{\lambda}\otimes \sigma_{\lambda}}\leq B^{\mathbf{x}\mathbf{y}}_{{\mathbf{a}\mathbf{b}}}$, then by mixing these inequality with probability distribution $p_{\lambda}$, we arrive at the conclusion. \qed
\end{pf}
\end{prop}

With the same spirit as the proof of proposition \ref{prop:f-e} we can prove the following proposition for steering detection. See Ref.~\cite{Pramanik2014} for the similar discussion in which they dealt with the problem of steering detection using some kinds of games.
\begin{prop}
If Alice and Bob share a state $\rho$ which admits LHS model for Bob's system, then for any observables $\mathbf{x}'$ and $\mathbf{x}$ chosen by Alice and Bob respectively, and for all outcome strings $\mathbf{a}'=(x'_1=a'_1,\cdots,x'_m=a'_m)$ of $\mathbf{x}'$ and $\mathbf{a}=(x_1=a_1,\cdots,x_m=a_m)$ of $\mathbf{x}$, the assemblage of Bob's system is $\mathscr{A}_{\mathbf{x}'}=\{\{\sigma_{a'_j|x'_j}:p(a'_j|x'_j)\}_{a'_j}\}_{x'_j}$, then we have
\begin{equation}\label{eq:f-steering}
\sum_{i=1}^m\sum_{a'_j} p(x_i)p(a'_j)p(a_i)_{\sigma_{a'_j|x'_j}}\leq B^{\mathbf{x}}_{\mathbf{a}},
\end{equation}
where $B^{\mathbf{x}}_{\mathbf{a}}$ is the uncertainty bound of $\mathbf{x}$ with outcome string $\mathbf{a}$.
\end{prop}

Note that Eqs. (\ref{eq:f-steering}) consist of $d^2\times m^2$ inequalities, violation of any one of them will lead the conclusion that the shared state $\rho_{AB}$ is steerable. If we choose mutually unbiased bases and let $p(x_i)=1/m$, then from Ref.~\cite{Rastegin2015}, we have steering test inequality as $\frac{1}{m}\sum_{i=1}^m p(a_i|a'_j)\leq \frac{1}{d}(1+\frac{d-1}{\sqrt{m}})$.

For example, in the most simple $2$-$2$-$2$ steering scenario, we can apply the fine-grained UR for $\mathbf{y}=( \sigma_x,\sigma_z)$ and $\mathbf{x}=(x_0,x_1)$, $\frac{1}{2}p(\pm|a_0)+\frac{1}{2}p(0/1|a_1)\leq \frac{1}{2}+\frac{1}{2\sqrt{2}}$, to detect steering. For a general two-qubit state $\rho_{AB}=\frac{1}{2^2}\sum_{\mu,\nu=0}^3 T^{\mu\nu}\sigma_{\mu}\otimes\sigma_{\nu}$, Alice choose to measure $x_i=\frac{1}{2}[\mathds{1}+(-1)^{a_i}\mathbf{s_i}\cdot \bm{\sigma}]=\sum_{\mu=0}^3 x_i^{\mu} \sigma_{\mu}$, where $a_i=0,1$. Since $|\pm\rangle \langle\pm|=\frac{1}{2}(\mathds{1}\pm \sigma_x)$ and $|0/1\rangle \langle 0/1|=\frac{1}{2}(\mathds{1} \pm \sigma_z)$ and using the formula $\mathrm{tr}(\sigma_{\mu}\otimes \sigma_{\nu}\rho_{AB})=T^{\mu \nu}$, we get $p(a_i)=\mathrm{tr}(x_i\otimes \mathds{1} \rho_{AB})=\sum_{\mu=0}^3 x_i^{\mu}T^{\mu 0}$ and $p(+,a_i)= \mathrm{tr}(x_i\otimes |+\rangle \langle+|\rho_{AB})=\frac{1}{4}+\sum_{j=1}^3 x^jT^{i1}$ and analogous for all other probabilities. Then we arrive at the following inequalities (for outcome string $(0, 0)$ of $\mathbf{x}$, the others can be obtained similarly):
\begin{align}\label{}
p(+|0_0)+p(0|0_1)\leq \frac{1}{2}+\frac{1}{2\sqrt{2}}, & \nonumber  \\
p(+|0_0)+p(1|0_1)\leq \frac{1}{2}+\frac{1}{2\sqrt{2}}, & \nonumber \\
p(-|0_0)+p(0|0_1)\leq \frac{1}{2}+\frac{1}{2\sqrt{2}},  & \nonumber \\
p(-|0_0)+p(1|0_1)\leq \frac{1}{2}+\frac{1}{2\sqrt{2}}.   &
\end{align}
These inequalities are the constraints of the correlation tensor $T^{\mu\nu}$, thus it can be well linked with the steering ellipsoids which focus on the correlation tensors~\cite{Jevtic2014}.
\subsection{Differences and connections with the universal UR based methods}
In general, fine-grained UR for a number of measurements $\mathbf{x}=\{x_i\}_{i=1}^m$ is completely different from universal UR, since, for fine-grained UR, we regard each probability distribution $\mathbf{p}_{x_i}(\rho)=(p(0|x_i),\cdots,p(d-1|x_i))$ as $d$ individual variables and we make some constraints by choosing one variable from each $\mathbf{p}_{x_i},\forall i=1,\cdots, m$, e.g., $p(a_{1}|x_1)+\cdots +p(a_{m}|x_m)\leq B_{\mathbf{a}}^{\mathbf{x}}$; but for universal UR, each distribution is always treated as a whole, we only make a constraint on the tensor product of these distributions, e.g., $\mathbf{p}_{x_1}(\rho)\otimes\cdots \otimes \mathbf{p}_{x_m}(\rho)\preceq \bm{\omega}_{\mathbf{x}}$. Thus detecting methods based on fine-grained UR and universal are complementary to each other.

Nevertheless, there are still some connections between them. To make this clear, consider two probability distributions $\mathbf{p}=(p_1,p_2)$ and $\mathbf{q}=(q_1,q_2)$ corresponding to two dichotomic observables $x$ and $y$ upon a state $\rho$ respectively, we assume that $p_1\geq p_2$ and $q_1\geq q_2$. We first note that the uncertainty bound $\bm{\omega}_{\mathbf{x}}$ of universal UR for $d$-outcome measurements has a very special form \cite{Friedland2013}:
\begin{equation}\label{eq:bound}
\bm{\omega}_{\mathbf{x}}=(\omega_1,\cdots,\omega_d,0,\cdots,0).
\end{equation}
Thus for two dichotomic measurements we have $\mathbf{p}\otimes \mathbf{q}\preceq \bm{\omega}_{x,y}=(\omega_1,\omega_2,0,0)$. If we choose Schur concave function $H_{\infty}$ as uncertainty quantifier, then we have $H_{\infty}(\mathbf{p}\otimes \mathbf{q})=H_{\infty}(\mathbf{p} )+H_{\infty}(\mathbf{q})=-(\log_2 p_1+\log_2 q_1)\geq -\log_2 \omega_1$, i.e., $\log_2 p_1+\log_2 q_1\leq \log_2 \omega_1$. This is, in some sense, a fine-grained UR, since it is just about some particular outcome string (here it is $(p_1,q_1)$ string) but not the whole distribution. But we still can not obtain the similar inequality for $(p_1,q_2)$, $(p_2,q_1)$ and $(p_2,q_2)$. By above construction, we can transform the detecting inequalities of entanglement and steering based on universal UR to some inequalities based on fine-grained UR.

\section{Conclusion and Discussion}
\label{sec:conclusion}
With the precise formulation of the $\mathcal{C}$-correlation criteria based on UR($\mathcal{C}$=entanglement, steering), we now briefly analyze the relationship between $\mathcal{C}$-correlations and uncertainty principle. Firstly,  pure entangled states must reveal Bell nonlocality \cite{Yu2012}, hence for pure states, entanglement implies steering. UR can detect all pure steerable states~\cite{Reid1989,Cavalcanti2009,Walborn2011}, therefore all pure entangled or steerable states can be detected by UR. And for many important families of states like two-qubit Werner states and isotropic states, our detection method can also give a very exact description of the entanglement range and steerable range by fine-grained UR with dichotomic measurements~\cite{Pramanik2014} and UR based on minimum entropy~\cite{Guhne2004a}.

In conclusion, universal UR provides us with a very general framework to detect entanglement and steering. The uncertainty quantifier $\Omega$ we choose is convex Schur concave function, thus many previous detection methods based on entropic UR are just some specified cases of our detecting methods. Fine-grained UR, in some sense, is a complementary part for universal UR, we develop the criteria based on fine-grained UR. Thus this work provides a very complete description of UR-based nonlocal correlation detections. Many unrevealed nonlocal states can be detected using our criteria. Our approach can be extended to the memory-assist scenario and genuine correlation detection, and as we have mentioned, there are some inherent difficulty in extending the method to detect Bell nonlocality, which are our future  projects.

\indent Z.-A. Jia acknowledges Alberto Riccardi for comments on example 1 and example 2 and acknowledges Chen Qian for many beneficial discussions. This work is supported by the National Natural Science Foundation of China (Grants No. 11275182 and No. 61435011), the National Key R \& D Program (Grant No. 2016YFA0301700), and the Strategic Priority Research Program of the Chinese Academy of Sciences (Grants No. XDB01030100 and No. XDB01030300).

\bibliographystyle{apsrev4-1}

\begin{thebibliography}{59}
\bibitem [{par()}]{partialorder}%
  \BibitemOpen
  \href@noop {} {\emph {\bibinfo {title} {In a strict sense, it is not a
  partial order, since it is not antisymmetic, i.e., we can not deduce
  $\mathbf{p}=\mathbf{q}$ from $\mathbf{p}\preceq\mathbf{q}$ and
  $\mathbf{q}\preceq\mathbf{p}$, we can only conclude that $\mathbf{p}=
  P\mathbf{q}$, where $P$ is a permutation matrix. But `$\preceq$' is still
  reflexive and transitive.}}}\BibitemShut {Stop}%
\bibitem [{ent()}]{ent}%
  \BibitemOpen
  \href@noop {} {\emph {\bibinfo {title} {Since we can add trailing zeros for
  outcomes whenever necessary, we still assume that each measurement has $d$
  outcomes. Using the same tools as in the proof of pososition 1, we can obtain
  the similar inequality $\sum_{i,j}\Omega(\mathbf{p}_{x_i\otimes y_j})\geq
  \max
  \{n\Omega(\bm{\omega}_{\mathbf{x}}),m\Omega(\bm{\omega}_{\mathbf{y}})\}$,
  note that $\bm{\omega}_{\mathbf{x}}$ is now deduced for $m$ measurements
  $\mathbf{x}$ but $\bm{\omega}_{\mathbf{y}}$ is deduced for $n$ measurements
  $\mathbf{y}$.}}}\BibitemShut {Stop}%
\end{thebibliography}
%

\end{document}